\documentclass[12pt]{iopart}

\usepackage{graphicx}

\begin{document}

\title{Complex and strongly anisotropic magnetism in the pure spin system EuRh$_2$Si$_2$}

\author{Silvia Seiro}
\address{Max Planck Institute for Chemical Physics of Solids, N{\"o}thnitzer Stra{\ss}e 40, 01187 Dresden, Germany}
 \ead{seiro@cpfs.mpg.de}
\author{Christoph Geibel}
\address{Max Planck Institute for Chemical Physics of Solids, N{\"o}thnitzer Stra{\ss}e 40, 01187 Dresden, Germany}

\date{\today}

\begin{abstract}

In divalent Eu systems, the $4f$ local moment has a  pure spin state J{=}S{=}7/2. Although the absence of orbital moment precludes crystal electric field effects, we report a sizeable magnetic anisotropy in single crystals of EuRh$_2$Si$_2$.  We observed a surprisingly complex magnetic behavior with three succesive phase transitions. The Eu$^{2+}$ moments order in a likely amplitude-modulated structure below 24.5\,K, undergoing a further transition to a structure that is possibly of the equal moment type, and a first order transition at lower temperatures, presumably into a spin spiral structure. The sharp metamagnetic transition observed at low fields applied perpendicular to the hard axis is consistent with a change from a spiral to a fan structure. These magnetic structures are presumably formed by ferromagnetic planes perpendicular to the \textbf{c} axis, stacked antiferromagnetically along \textbf{c} but not of type I, at least just below the ordering temperature.  Since EuRh$_2$Si$_2$ is isoelectronic to EuFe$_2$As$_2$, our results are also relevant for the complex Eu-magnetism observed there, especially for the transition from an antiferromagnetic to a ferromagnetic state observed in EuFe$_2$P$_2$ upon substituting As by P.
\end{abstract}

\submitto{\JPCM}
\maketitle

\section{Introduction}

Competing interactions generally give rise to complex behavior. This is particularly true for magnetic systems, where the interplay of magnetic interactions between nearest, next nearest, and further neighbors can lead to complex magnetic ordering and frustration. In the case of rare earth intermetallic compounds, magnetic exchange between localized rare earth moments is mediated by conduction electrons, giving rise to long range interactions of oscillatory character. In addition,  in most rare earth compounds the magnetism is strongly influenced by the single ion anisotropy due to the crystal electric field, notably manifesting in the presence of metamagnetic transitions. Systems based on trivalent Gd and divalent Eu, with a pure spin configuration $\rm S{=}7/2$ and $\rm L{=}0$, provide the unique opportunity of studying the effect of exchange in the absence of single-ion anisotropy. Since spherically symmetric, such a configuration is inert to the crystal electric field and compounds based on Gd$^{3+}$ or Eu$^{2+}$ are expected to have at most a small magnetic anisotropy.  EuRh$_2$Si$_2$, the Eu homologue of the paradigmatic heavy fermion YbRh$_2$Si$_2$~\cite{Trovarelli00} and the Kondo antiferromagnet CeRh$_2$Si$_2$~\cite{Severing89},  is a metal which orders antiferromagnetically below T$_{\rm N}\sim25$\,K \cite{Hossain01}. So far, little is known about the magnetic structure of EuRh$_2$Si$_2$. Early M\"{o}ssbauer spectroscopy measurements~\cite{Felner84} on polycrystals determined that the Eu spins lie in the basal plane at 4.1\,K but neither the magnetic propagation vector nor the orientation close to T$_{\rm N}$ of the ordered moments  have been determined. More recent work on polycrystalline samples revealed features at roughly 23 and 25\,K in magnetization and specific heat, and metamagnetism at low temperatures~\cite{Hossain01}.  Here, we report the physical properties of  EuRh$_2$Si$_2$ single crystals. This compound presents a sizeable magnetic anisotropy both in the paramagnetic and the ordered phases, in spite of the absence of crystal field effects. We show that in EuRh$_2$Si$_2$ different magnetic structures are close in energy, giving rise to complex magnetism. We identify a third phase transition occurring at $\sim 14$\,K, likely into a spin spiral magnetic structure, which undergoes a metamagnetic transition at low fields possibly into a fan structure.

\section{Experimental Details}

The EuRh$_2$Si$_2$ single crystals were synthesized as described in Ref.~\cite{Seiro10}. The structure was characterized by powder x-ray diffraction on a crushed crystal, using Cu K$\alpha$ radiation and LaB$_6$ ($a=4.1569$\,\AA) as an internal standard.
The chemical composition was checked by energy-dispersive x-ray
spectroscopy (EDX). Heat capacity and four-point resistivity
measurements were performed in a Quantum Design PPMS system. Magnetization data were taken using a Quantum Design MPMS
SQUID magnetometer.

\section{Results and Discussion}

The chemical composition determined by EDX microprobe analysis was ($20\pm1$)\,at\%  Eu, ($40\pm2$)\,at\%  Rh and ($40\pm2$)\,at\% Si. 
X-ray diffraction measurements confirmed a tetragonal structure for the crystals, with
lattice parameters a${=}(4.0892\pm0.0002)$\,\AA, and
c${=}(10.219\pm0.0006)$\,\AA, in agreement with the values
published for polycrystalline samples~\cite{Chevalier86,Hossain01}.
No secondary phases were detected. Laue images show that the
crystals grow preferentially along the basal plane, with the
shortest dimension (typically 500-900\,$\mu$m) along the \textbf{c} axis.

\begin{figure}
\includegraphics[width=0.7\textwidth]{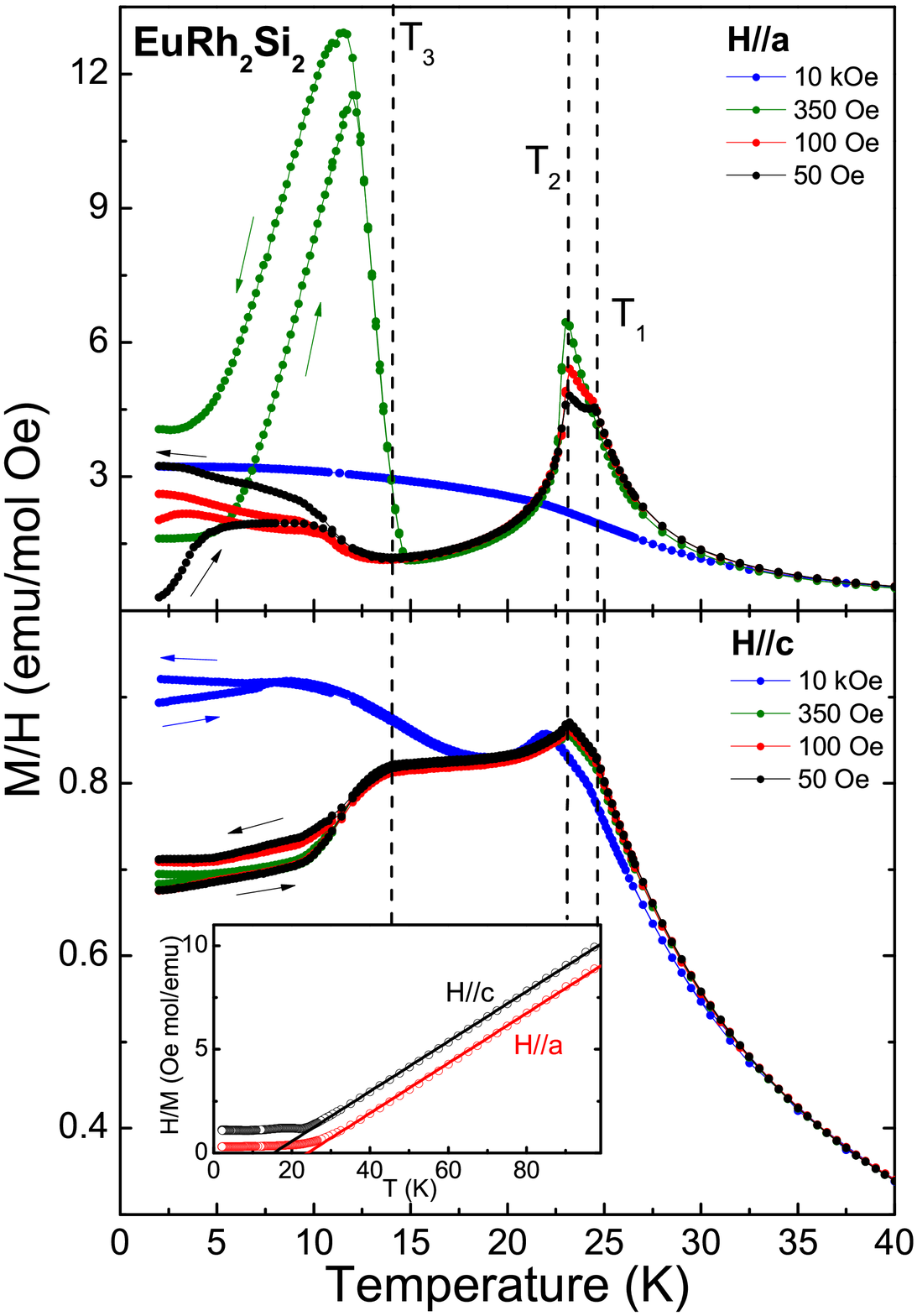}
\caption{\label{MvsT} Magnetic susceptibility as a function of
temperature   for H${\parallel}\bf{a}$ (top panel) and  H${\parallel}\bf{c}$ (bottom panel). Data were taken upon warming from a zero-field-cooled initial state and upon cooling in an applied field, as indicated by the arrows. Inset: Inverse susceptibility vs. temperature. The solid lines represent fits to the data with a Curie-Weiss law.}
\end{figure}

Magnetic susceptibility data as a function of temperature in different applied fields (H) are shown in Figure~\ref{MvsT} for H${\parallel}\bf{a}$ and  H${\parallel}\bf{c}$. There is a sizeable anisotropy, susceptibility for  H${\parallel}\bf{a}$ being larger than for  H${\parallel}\bf{c}$.  The anisotropy at low fields cannot be explained by the sample shape although the platelets are wide and thin. The susceptibility ratio between the two field orientations is 1.6 at 40\,K and 5.3 at 23\,K. Above $\sim 35$\,K susceptibility follows a Curie-Weiss behavior both for  H${\parallel}\bf{a}$ as well as H${\parallel}\bf{c}$.  The effective magnetic moment $\mu_{\rm{eff}}{=}(8.2\pm0.2)$\,$\mu_{\rm B}$ is the same for both field directions and close in value to that expected for Eu$^{2+}$ (7.94\,$\mu_{\rm B}$). The Curie-Weiss temperature $\Theta$ is positive for both field orientations indicating a dominant ferromagnetic exchange, with $\Theta_{\rm a}{=}(24.0\pm0.05)$\,K somewhat larger than $\Theta_{\rm c}{=}(15.3\pm0.3)$\,K but both nevertheless close in absolute value to the ordering temperature. Since for Eu$^{2+}$ the magnetic moment has purely spin character and no crystal field effects are expected, the large anisotropy observed in the susceptibility close to T$_{\rm N}$ results from a rather moderate exchange anisotropy ($(\Theta_{\rm a}-\Theta_{\rm c})/\Theta_{\rm a}\sim 36$\,\%); this effect is enhanced by the divergence of the susceptibility upon approaching the Curie-Weiss temperature due to the global ferromagnetic nature of the exchange. Since $4f$ electrons are well localized, the magnetic coupling must be indirect through the conduction electrons (RKKY). This also explains the difference in ordering temperature with its homologue GdRh$_2$Si$_2$ (T$_{\rm N}{=} 108$\,K)~\cite{Szytula92}, since the different valence states of the two rare earths imply an additional $d$ electron in the conduction band for Gd$^{3+}$. In addition, the anisotropy of the exchange in EuRh$_2$Si$_2$ implies some spin-orbit effects on the electronic states close to the Fermi level.

In a simple molecular field picture, a ferromagnetic Curie-Weiss temperature very close in absolute value to the actual antiferromagnetic ordering temperature points out to a strong coupling within each ferromagnetic sublattice and a much weaker intersublattice coupling. A comparison across the rare earth (R) series of the magnetic properties of RRh$_2$Si$_2$ compounds, shows that in all reported cases there is a transition to an antiferromagnetically ordered state at a temperature T$_{\rm N}$ that scales roughly with the de Gennes expression $(g_J-1)^2 J(J+1)$ for the heavy rare earths (Gd to Yb) and departs considerably from it for the light rare earths (Ce to Eu)~\cite{Szytula08}. More importantly, the magnetic structure of RRh$_2$Si$_2$ has been determined for most members of the series~\cite{Szytula08,Quezel84, Welter03, Melamud84, Slaski83,Szytula84,Yakinthos86} and has been found to be predominantly of type I, with rare earth moments ordering ferromagnetically in the basal plane and stacking antiferromagnetically along \textbf{c}, i.e. with a propagation vector $\tau{=} (0,0,1)$. This ubiquity of ferromagnetic planes antiferromagnetically stacked along \textbf{c} suggests a similar kind of ordering for EuRh$_2$Si$_2$. 

Upon cooling below 25\,K, a double feature is observed at T$_1$=24.5\,K and T$_2$=23.2\,K  in the susceptibility at low fields, below which the susceptibility decreases rapidly for H${\parallel}\bf{a}$  and much more gradually for H${\parallel}\bf{c}$.The characteristic temperatures of these anomalies are only weakly dependent on H${\parallel}\bf{c}$, but for H${\parallel}\bf{a}$ the feature at T$_1$ is rapidly suppressed, merging with that at T$_2$ already in a field of 350\,Oe.  All this hints towards an antiferromagnetic order with an easy plane perpendicular to the \textbf{c} axis. Indeed, early M\"{o}ssbauer experiments performed on EuRh$_2$Si$_2$ determined that the moment lies in the basal plane at 4.2 K~\cite{Felner84}.  

\begin{figure}
\includegraphics[width=0.7\textwidth]{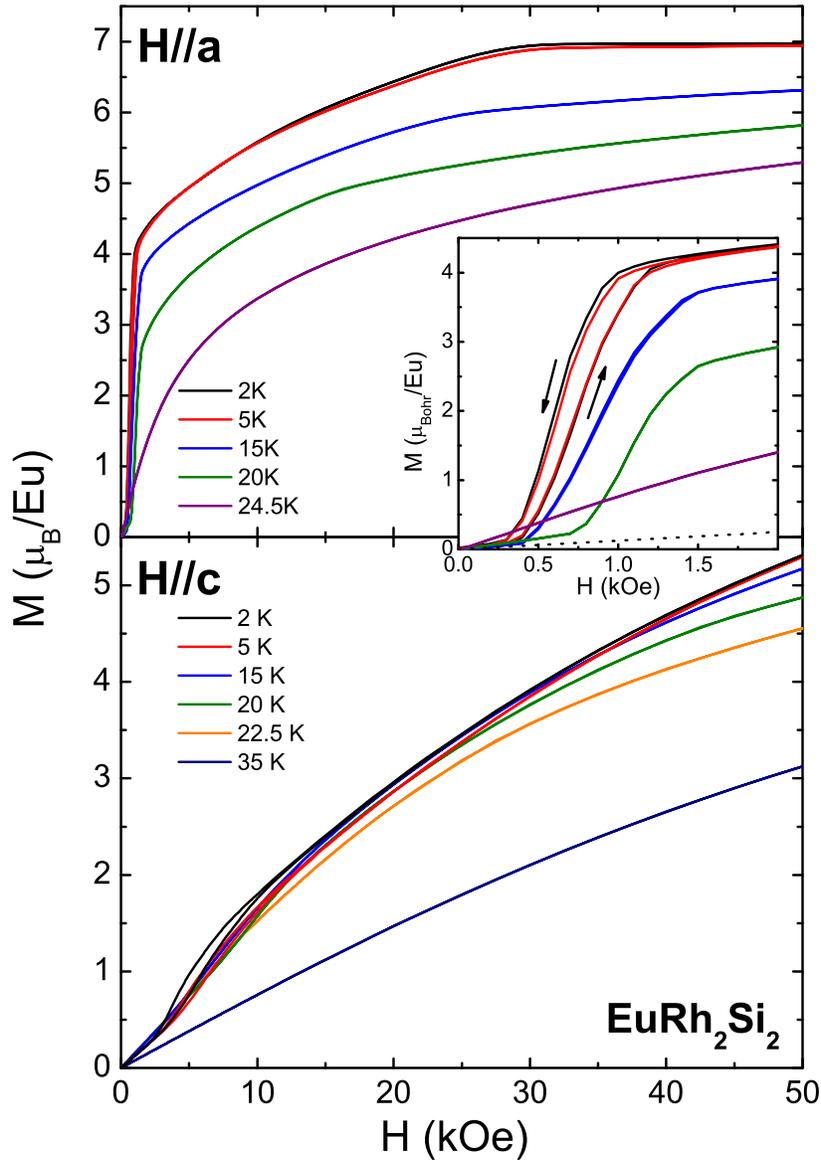}
\caption{\label{MvsH} Magnetization curves  measured from a zero-field-cooled initial state as a function of field applied along \textbf{a} (top panel) and \textbf{c} (bottom panel) at different temperatures. The inset shows a blowup of the top panel for low fields; the 2\,K curve for H${\parallel}\bf{c}$ is also plotted for comparison (dotted line). The arrows indicate the direction of the field sweep.}
\end{figure}

Below  $\rm{T}_3\sim14$\,K a further anomaly is observed. Susceptibility increases upon cooling for  H${\parallel}\bf{a}$ and drops strongly for H${\parallel}\bf{c}$. For lower temperatures, an irreversibility  between zero-field-cooled and field-cooled curves appears, indicating the presence of a ferromagnetic component. Comparison with magnetization curves as a function of applied field in Figure~\ref{MvsH} reveals that at the lowest fields (H$\leq100$\,Oe) the irreversibility is most probably a parasitic effect, while it is related to the presence of a metamagnetic transition at somewhat higher fields.

The presence of magnetic anisotropy is also evident from magnetization curves in Figure~\ref{MvsH}. For H$\parallel\bf{a}$, a step-like hysteretic increase of roughly 4$\mu_{\rm B}$ in magnetization occurs at low temperatures and fields, beyond which magnetization increases continuously with field, reaching the saturation moment of 7$\mu_{\rm B}$ per Eu atom at $\sim 30$\,kOe. Although such behavior is reminiscent of the spin flop transition occurring in antiferromagnets with weak magnetocrystalline anisotropy when field is applied in the easy direction, it is striking to observe it in this Eu system where no crystal field effects are present, so that the magnetization is expected to evolve continuosly as a function of  H${\parallel}\bf{a}$. The abrupt  increase observed in magnetization suggests rather a flipping of the moments of whole ferromagnetic planes, analogous to the case of TbRh$_2$Si$_2$.  In the latter, which at zero field presents a type I magnetic structure (i.e. a stacking   $+-+-$ of ferromagnetic planes along the easy axis \textbf{c}), a field of $\sim 9$\,T parallel to  \textbf{c} induces an abrupt transition to a stacking of the form $+-++$~\cite{Himori07}. Therefore, one could think that the abrupt metamagnetic transition observed in EuRh$_2$Si$_2$ occurs when the energy barrier separating antiferromagnetic and ferri- or ferromagnetic configurations is suddenly overcome by the magnetic field. For example, a flipping of every fourth layer would result in a magnetization change of 3.5$\mu_{\rm B}$. However, in contrast to TbRh$_2$Si$_2$, which is a system with a strong Ising character,  EuRh$_2$Si$_2$ is an easy plane system where hardly any differences are observed when applying the field along [1\,1\,0] instead of [1\,0\,0]. It seems then unlikely that it should take a finite field in the easy plane to flip the orientation of a ferromagnetic layer.  Another source of sharp metamagnetism is the transition from a spin spiral (or screw) structure to a fan (or sinusoid) structure, as occurs in MnP: the application of a field perpendicular to the hard direction \textbf{a} induces a first order transition from a spiral magnetic structure with propagation vector along \textbf{a} and moments in the \textbf{bc} plane to a fan structure with propagation vector along \textbf{a} but where the moments oscillate in the \textbf{bc} plane around the field direction~\cite{Nagamiya71,Becerra76}.  In any case, the energy barrier separating both spin structures in EuRh$_2$Si$_2$ is very small ($\Delta\mu~{\rm H}/k_{\rm B}< 1$\,K).

In the case of H${\parallel}\bf{c}$, the hard magnetic axis, the magnetization increases much more slowly upon increasing field, and the full Eu moment is not recovered up to 50\,kOe. A small broad hysteretic magnetization jump  is visible at low temperatures for fields above 2500\,Oe, but it is not clear whether this feature is intrinsic to the magnetism in the hard direction or due to a small field component along the easy plane originating from a slight sample misalignment.

\begin{figure}
\includegraphics[width=1.0\textwidth]{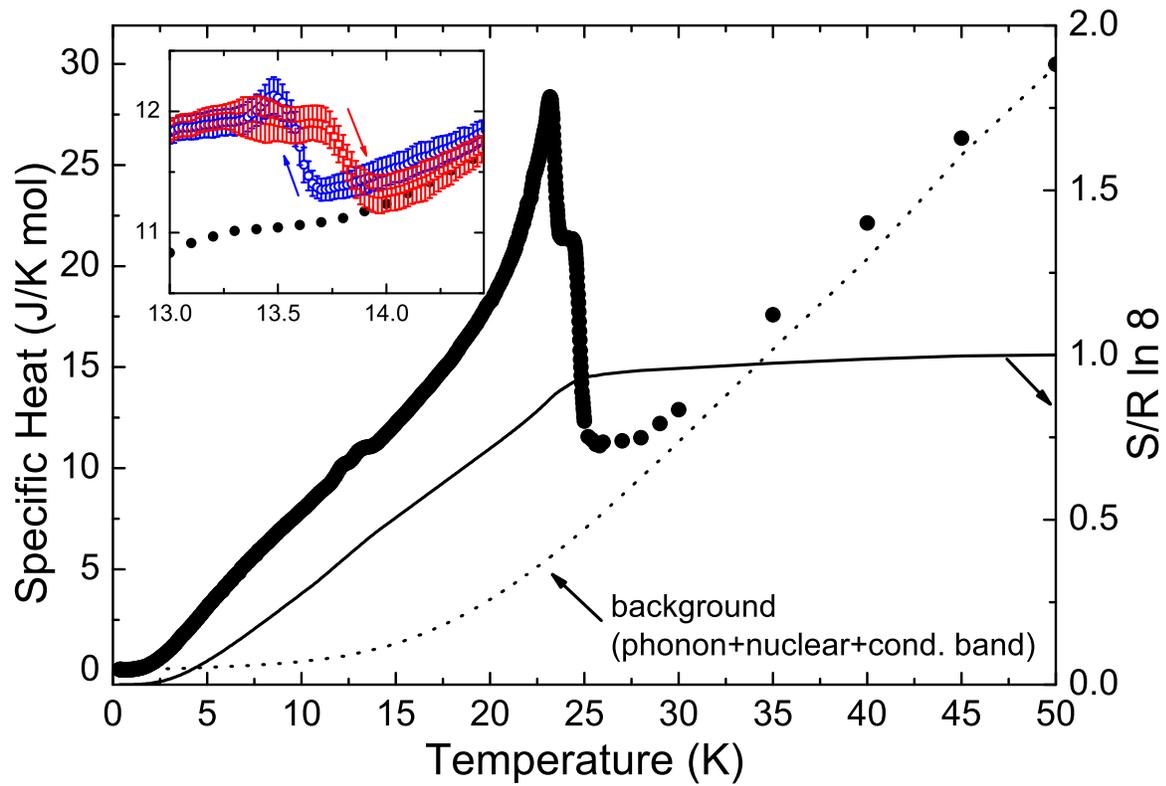}
\caption{\label{Figure2} Specific heat (left axis) and entropy (right axis) as a function of
temperature for EuRh$_2$Si$_2$. The background due to phonons, nuclear moments and electrons is represented by a dotted line (see text). Inset: Specific heat data around T$_3$ obtained using the PPMS standard (black circles) and the single slope analysis for warming (red open squares) and cooling (blue open circles)  (see text).}
\end{figure}

The specific heat as a function of temperature in Figure~\ref{Figure2}  exhibits anomalies at 24.8\,K and 23.5\,K, consistent with those observed in the magnetic susceptibility at T$_1$ and T$_2$. A small broad anomaly is also detected close to T$_3$.  It must be noticed that the standard evaluation of the specific heat performed by PPMS (black circles in Figure~\ref{Figure2}) assumes that the sample specific heat remains constant during and after the application of the heat pulse.  If the specific heat of the sample varies strongly over a small temperature range, as in a first order transition, the values extracted by the standard PPMS evaluation procedure are not reliable. We have therefore performed a slope analysis of the relaxation data in the vicinity of T$_3$~\cite{PPMSmanual}, using either the warming or cooling part of the curve. The results are plotted in the inset of Figure~\ref{Figure2}: a small but clear specific heat jump is revealed at T$_3$, which presents a small hysteresis (0.2\,K) between warming and cooling data. These features are robust,  unlike the oscillations observed in the lower temperature range which are affected by the analysis parameters (averaging width and excluded initial region). This indicates that a thermodynamic first-order phase transition occurs at T$_3$. 

In isotropic systems without single ion anisotropy, the evolution of the specific heat as a function of temperature and in particular the specific heat jump at T$_{\rm{N}}$ can provide information about the type of magnetic order~\cite{Blanco91}. The specific heat jump at T$_1$ is similar in magnitude  to the value proposed for a (collinear) amplitude modulated structure (13.43 J/mol K)~\cite{Blanco91}, ruling out simple antiferromagnetic, helical and cycloidal equal-moment structures just below T$_1$.  It must be noticed that if anisotropic exchange is explicitly considered, non-collinear amplitude-modulated structures can form~\cite{Rotter01}, presenting a somewhat higher specific heat jump at T$_{\rm N}$.  However, the fact that it is very difficult to align the moments to a field parallel to the \textbf{c} axis suggests the moments are confined to the basal plane.  Such amplitude-modulated magnetic structures usually undergo for entropic reasons a first order transition upon cooling into an equal-moment structure which can have a different propagation vector, or evolve into a squared-up arrangement with the development of components at higher order harmonics of the propagation vector~\cite{Blanco91EPL}. Since T$_1$ and T$_2$ are very close, in a first approximation the sum of the specific heat jumps at T$_1$ and T$_2$  should be similar to the magnetic specific heat change that would be expected for a single transition occuring at T$_2$. This value is close to that expected for equal-moment structures, suggesting the transition at T$_2$ is to an equal-moment state. On the other hand, the transition at T$_3$ is first-order in nature, hinting to a different symmetry of the magnetic states above and below T$_3$. At low temperatures, the characteristic broad Schottky-like bump~\cite{Blanco91EPL} present in 4$f^7$ magnetic systems at T$_{\rm{N}}/4$  can be observed in the specific heat.

Since no specific heat data are available for A$^{2+}$Rh$_2$Si$_2$, the phonon contribution was estimated from the specific heat data of isostructural Lu$^{3+}$Rh$_2$Si$_2$, from which an electronic contribution with a Sommerfeld coefficient of 7 mJ/K$^2$mol was subtracted~\cite{Ferstl07}.  The Debye temperature for the Lu compound is expected to be higher because of the additional binding provided by the $5d$ electron.   In order to account for the change in the characteristic Debye temperature due to the divalent configuration of Eu in EuRh$_2$Si$_2$, a rescaling of the Debye temperature $\Theta_{\rm D}^{\rm Eu}/\Theta_{\rm D}^{\rm Lu}{\sim}0.965$ was considered in the estimation of the phonon contribution,  so that the magnetic entropy Rln8 expected for Eu$^{2+}$ atoms with a J{=}7/2 moment is fully recovered at T${\sim} 2 \rm{T}_1$, where the contribution of magnetic fluctuations is expected to be negligible. A nuclear contribution (evident at very low T, see Ref.~\cite{Seiro10}) and a conduction band contribution estimated to 25 mJ/K$^2$ ~\cite{Seiro10} were subtracted in order to obtain the magnetic specific heat. The magnetic entropy estimated in this manner
 reaches 0.93\,Rln8 at T$_1$, see Figure~\ref{Figure2}. The missing entropy (7\,\% of Rln8) is less than expected from numerical calculations of simple magnetic lattices (14-20\,\%), indicating that short-range fluctuations above \,T$_N$ are comparatively weak~\cite{Mattis85}.

\begin{figure}
\includegraphics[width=1.0\textwidth]{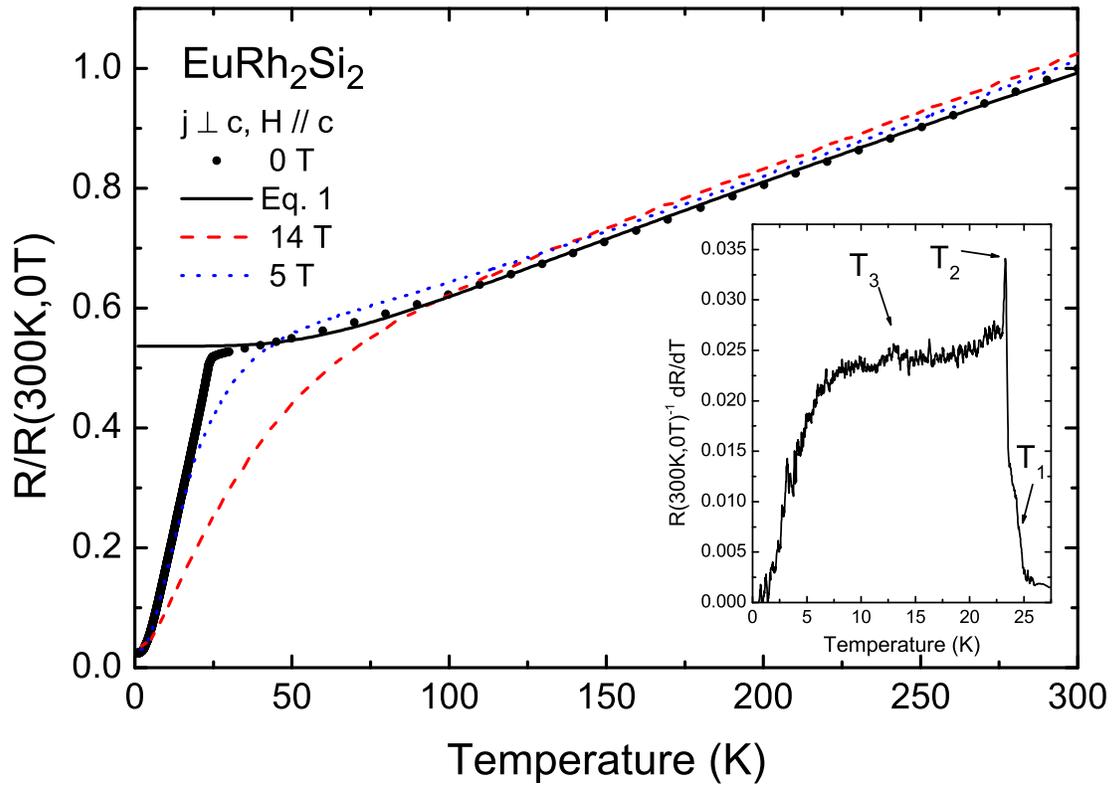}
\caption{\label{Resistivity}  Resistivity ratio measured as a
function of temperature for current flowing along the [1\,1\,0] direction in magnetic fields of 0, 5 and 14\,T applied along [0\,0\,1]. The black continuous line represents a fit with Eq.~\ref{Eq1} (see text).  Inset: Derivative of the zero-field resistivity ratio as a function of temperature showing anomalies corresponding to the magnetic transitions.}
\end{figure}

Figure~\ref{Resistivity} shows the temperature dependence of the resistivity $\rho$ normalized to its zero-field value at room temperature. Upon cooling from room temperature,  the resistivity first decreases roughly linearly but then gradually reduces its slope, as is characteristic for the phonon scattering contribution in a metal. Below T$_1$ the resistivity drops more rapidly on entering the magnetically ordered phase, consistent with the reduction in spin-disorder scattering, and below T$_2$, a stronger suppression of the resistivity takes place with $d\rho/dT$ being one order of magnitude larger than in the paramagnetic phase. For the homologue compound GdRh$_2$Si$_2$, the resistivity slope increases only a factor of $\sim 2$ below T$_{\rm N}$~\cite{Duijn00}. This indicates that spin-flip scattering is particularly strong in EuRh$_2$Si$_2$. A small broad anomaly in $d\rho/dT$ could be distinguished at T$_3$. The residual resistivity ratio $\rho(300K)/\rho(0.5 K)\sim40$ indicates the high quality of the crystals and corresponds to a residual resistivity is of the order of 1\,$\mu\Omega$~cm. The zero-field resistivity data can be fitted above 30\,K with the expression
 
\begin{equation} \label{Eq1}
\rho(T) = \rho_{c} + \rho_{BG} \left(\frac{T}{ \Theta_R}\right)^5 \int_0^{\Theta_R/T} \frac{x^5}{ (e^x -1) (1-e^{-x})} dx,
\end{equation}

\noindent where the constant $\rho_{c}=\rho_{0}+\rho_{sd}$, with $\rho_0$ the residual resistivity and $\rho_{sd}$ the spin-disorder contribution, while the second term corresponds to the phonon contribution $\rho_{ph}(T)$ given by the Bloch-Gr\"uneisen expression. This yields a Debye temperature $\Theta_{\rm R}{=}377$\,K and a spin-disorder contribution  $\rho_{sd}{=}3371 \rho_{ph}(\rm{T_N})$. Applying a magnetic field along the \textbf{c} axis results in a positive transverse magnetoresistance at high temperatures, which is at 14\,T roughly a factor 4 larger than for polycrystalline copper~\cite{deLaunay59}. The magnetoresistance (MR) becomes large and negative upon cooling towards the N\'{e}el temperature,  reaching 53\%  at 14\,T. Such an important negative magnetoresistance at temperatures around the onset of magnetic order is usually observed for ferromagnets (for example  Ni~\cite{Belov94}, GdNi~\cite{Mallik97}, GdAl$_2$~\cite{Lang00}, Gd$_2$Co$_2$Ga~\cite{Sengupta05prb}, Gd$_2$Co$_2$Al~\cite{Sengupta05prb}, Eu$_2$CuSi$_3$~\cite{Majumdar99}, EuFe$_2$P$_2$~\cite{Feng10}). In antiferromagnets, on the other hand, a large negative magnetoresistance around T$_{\rm N}$ is not systematically observed: while some compounds exhibit a small positive MR only below T$_N$ (e.g. GdCu$_6$~\cite{Chattopadhyay12}, GdCu$_2$Ge$_2$~\cite{Mallik98}), others show practically no MR (as GdAu$_2$Si$_2$~\cite{Mallik98} or GdPd$_2$Ge$_2$~\cite{Mallik98}) and a few (GdPtP~\cite{Lang00}, Eu$_{14}$MnBi$_{11}$~\cite{Chan98}, Eu$_3$Ni$_4$Ga$_4$~\cite{Anupam12}, Gd$_7$Rh$_3$~\cite{Sengupta05}) have a large negative MR around T$_{\rm{N}}$. The latter show often predominantly ferromagnetic interactions, manifest for example as a positive Curie-Weiss temperature in the paramagnetic susceptibility, as is also the case for EuRh$_2$Si$_2$. 

\begin{figure}
\includegraphics[width=1.0\textwidth]{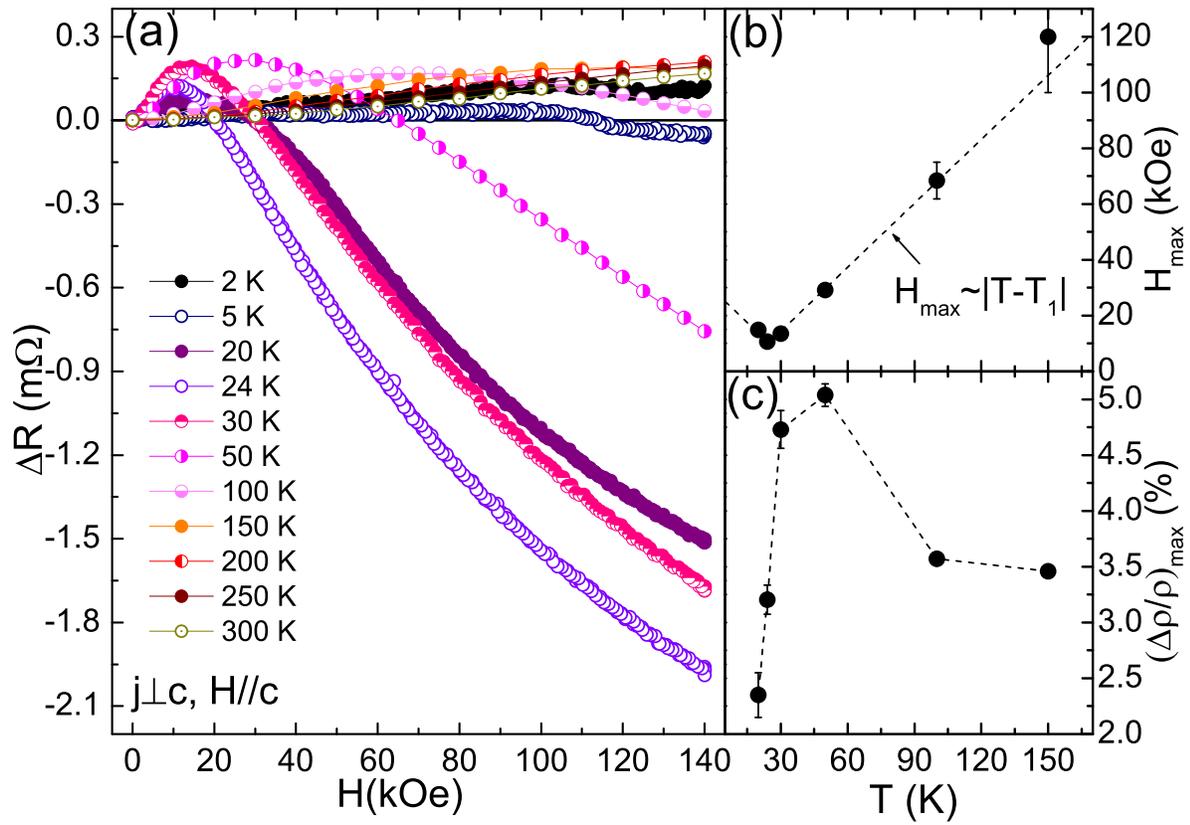}
\caption{\label{mrsummary}(a) Resistance change relative to zero field as a function of magnetic field applied along the c axis for different temperatures. (b) Field at which magnetoresistance shows a maximum. (c) Temperature evolution of the magnetoresistance maximum.}
\end{figure}

At low fields, the transverse magnetoresistance measured for H${\parallel}\bf{c}$ is positive for all measured temperatures, see Figure~\ref{mrsummary}. For both isotherms measured below T$_3$, a steplike decrease of MR takes place at $\sim 110$ kOe, suggesting the presence of a metamagnetic transition. A comparison with magnetization data (which are only available up to 50 kOe) suggests that this critical field might correspond to the saturation field. Well above T$_3$, at 20\,K, the steplike decrease is no longer present, but a broad maximum is observed beyond which the magnetoresistance becomes increasingly negative on increasing field. No substantial change in the shape of the MR curves is observed by heating through T$_2$ and T$_1$ into the paramagnetic phase, although the position of the broad maximum varies strongly with temperature. In a relaxation time approximation, the magnetoresistance follows a universal curve as a function of $\omega_c \tau$, where the cyclotron frequency $\omega_c{=}q\rm{ H}/m^*$.  In  "normal" compensated metals the magnetoresistance in the classical regime is always positive because the scattering process itself is not field dependent: the only effect of the field is to modify the electron orbit, thus giving rise to Kohler scaling $\rho/\rho_0 \propto f(\rm{H}/\rho_0)$. In EuRh$_2$Si$_2$ Kohler scaling breaks down: the broad MR peak moves to fields that are at least a factor of two larger between 20 and 50\,K, while $\rho_0$ changes very little in the same temperature range, see Figure~\ref{Resistivity}. Figure~\ref{mrsummary}  rather suggests a scaling of the form $\Delta\rho/\rho_{max} \propto f(\rm{H/|T{-}T_1|})$.   A relaxation time $\tau \propto 1/|\rm{T{-}T_1}|$ is suggestive of spin-flip scattering. The negative sign accompanying T$_1$ can easily be accounted for by the ferromagnetic exchange. This exchange supports the external field in polarizing the local moments and thus in suppressing the spin-flip scattering. On the other hand we notice that at very low (2\,K) and very high (200-300\,K) temperatures, where field induces changes on the spin-flip scattering mechanism are weak, the magnetoresistance is positive, almost linear in field (at least below 100\,kOe) and of roughly the same absolute size. Thus, the magnetoresistance in EuRh$_2$Si$_2$ seems to be essentially a sum of a linear in H "normal" metallic contribution and a negative spin-flip related magnetic contribution which scales as H/(T-T$_1$). However, an additional effect is needed to explain an enhancement of the positive contribution at low fields in the vicinity of the N\'eel temperature.  It is worth mentioning that a similar scaling of MR $\propto 1/(\rm{T{-}T_C})$  was observed in the ferromagnet EuFe$_2$P$_2$  but erroneously taken as evidence for Kondo interactions~\cite{Feng10}.

In summary, single crystals of EuRh$_2$Si$_2$ exhibit a marked magnetic anisotropy with the easy plane perpendicular to the \textbf{c} in spite of the pure spin character of Eu$^{2+}$. The positive Curie-Weiss temperature close to T$_{\rm N}$  in magnitude, together with the comparison to other RRh$_2$Si$_2$ systems, suggests strong ferromagnetic coupling within the basal plane, and an antiferromagnetic stacking of these planes along \textbf{c}. Three succesive magnetic transitions occur upon cooling in zero field: A transition from the paramagnetic to an amplitude-modulated state at T$_1{=}24.5$\,K, followed closely by a transition at T$_2{=}23.8$\,K presumably into an equal-moment magnetic structure, and a third transition which is of the first order into a magnetic state which may be a spin spiral. The sharp metamagnetic transition observed as a function of field at low temperatures would then be related to a transition from spiral to fan magnetic structure, or be the signature of the collective flipping of some ferromagnetic moments.  This indicates that in EuRh$_2$Si$_2$ several magnetic configurations are very close in energy.  Our results may also be relevant to the complex Eu magnetism observed in the isoelectronic system EuFe$_2$(As$_{1-x}$P$_{x}$)$_2$, especially for the transition from  antiferromagnetism to ferromagnetism on increasing $x$.

\section*{Acknowledgements}
S. S. gratefully acknowledges discussions with Dr. R. Cardoso-Gil and Dr. M. Rotter.

\end{document}